\documentclass{PoS}

\usepackage[utf8]{inputenc}
\usepackage{caption}
\usepackage{subcaption}
\usepackage{lineno}

\usepackage{mathrsfs}
\usepackage[intlimits,centertags]{amsmath}
\usepackage{amssymb,amsfonts}
\usepackage{enumerate}
\usepackage{floatrow}
\usepackage{graphicx}
\usepackage{booktabs}

\title{Combined Analysis of Cosmic-Ray Anisotropy with IceCube and HAWC}

\ShortTitle{Combined Cosmic-Ray Anisotropy with IceCube and HAWC}

\author{The HAWC Collaboration$^{1}$, The IceCube Collaboration$^{2}$   
\\
{\it 
$^1$ \href{http://www.hawc-observatory.org/collaboration/icrc2017.php}{http://www.hawc-observatory.org/collaboration/icrc2017.php} \\
$^2$ \href{http://icecube.wisc.edu/collaboration/authors/icrc17_icecube}{http://icecube.wisc.edu/collaboration/authors/icrc17\_icecube}\\
}

E-mail: \email{juan.diazvelez@alumnos.udg.mx}
}

\abstract{During the past two decades, experiments in both the northern and southern hemispheres have observed a small but measurable energy-dependent sidereal anisotropy in the arrival direction distribution of Galactic cosmic rays with relative intensities at the level of one per mille. Individually, these measurements are restricted by limited sky coverage, and so the power spectrum of the anisotropy obtained from any one measurement displays a systematic correlation between different multipole modes $C_\ell$. We present the results of a joint analysis of the anisotropy on all angular scales using 
cosmic-ray data collected during 336 days of operation of the High-Altitude Water Cherenkov (HAWC) Observatory (located at 19$^\circ$ N) and 5 years of data taking from the IceCube Neutrino Observatory (located at 90$^\circ$ S) The results include a combined sky map and an all-sky power spectrum in the overlapping energy range of the two experiments at around 10 TeV. We describe the methods used to combine the IceCube and HAWC data, address the individual detector systematics, and study the region of overlapping field of view between the two observatories.

\vspace{4mm}

{\bf Corresponding authors:}
\speaker{J. ~C. D\'iaz-V\'elez$^{a,b}$},
M. Ahlers$^{c}$,
P. Desiati$^{b}$,
D. Fiorino$^{d}$,
\\
     \llap{$^a$}Centro Universitario de los Valles, Universidad de Guadalajara, Guadalajara, Jalisco, M\'exico \\
     \llap{$^b$}Wisconsin IceCube Particle Astrophysics Center (WIPAC) and Department of Physics, 
     University of Wisconsin--Madison, Madison, WI 53706, USA \\
      \llap{$^c$}Niels Bohr Institute, University of Copenhagen, Copenhagen, Denmark \\
     \llap{$^d$}Department of Physics, University of Maryland, College Park, MD, USA 
     
}

\FullConference{35th International Cosmic Ray Conference - ICRC217 -\\
		10-20 July, 2017\\
		Bexco, Busan, Korea}

\begin{document}

\section{Introduction}
Over the last few decades, several studies have measured
a small but significant variation in the intensity of cosmic rays of
medium and high energies as a function of right ascension.
This anisotropy was observed at energies of order 1 TeV and higher by a number 
of experiments including Tibet AS$\gamma$ \cite{Tibet:2005jun}, Super-Kamiokande \cite{SuperK:2007mar}, 
Milagro \cite{Milagro:2008nov}, EAS-TOP \cite{Aglietta:2009feb}, MINOS \cite{MINOS:2011icrc}, ARGO-YBJ \cite{ARGO:2013jun}, 
and HAWC \cite{HAWC:2014dec} in the Northern Hemisphere
and IceCube \cite{IceCube:detector, IceCube:2010aug,IceCube:2011oct,IceCube:2012feb} and its surface air shower array
IceTop \cite{IceCube:2013mar} in the Southern Hemisphere.
In both hemispheres, the observed anisotropy has two main features: a large-scale structure with an amplitude of about 
$10^{-3}$,  and a small-scale structure with an amplitude of $10^{-4}$  and angular size from $10^\circ$ to $30^\circ$. 

A number of theories have proposed scenarios where the large-scale anisotropy results from the distribution of cosmic ray sources in the Galaxy and of their diffusive propagation 
\cite{SuperK:2007mar, Milagro:2008nov, HAWC:2014dec, IceCube:2011oct, IceCube:2012feb,Tibet:2006oct, Milagro:2009jun, 
Amenomori:2012uda, 
ARGO:2013oct,Bartoli:2015ysa,
Erlykin:2006apr, Blasi:2012jan, Ptuskin:2012dec, Pohl:2013mar, 
Sveshnikova:2013dec, Kumar:2014apr, Mertsch:2015jan}. 
However, the origin of the small-scale anisotropy is less well understood since it
is expected that cosmic rays should lose any correlation with their original direction due to
diffusion as they traverse through interstellar magnetic fields.
There are several theories regarding the origin of this anisotropy, including structures in the heliomagnetic field, non-diffusive propagation, and turbulence in Galactic magnetic fields; see review in \cite{Ahlers:2016rox}.

Analyses of data from Earth-based experiments with partial sky coverage suffer from systematic effects and statistical uncertainties 
of the calculated angular power spectrum. In this analysis we combine data from 
IceCube and HAWC at the same energy to study the full-sky anisotropy.
Important information can be obtained from the power spectrum at low-$\ell$ (large scale), 
which is the region most affected by partial sky-coverage of one experiment only.
However, it should be noted that neither observatory is sensitive to variations across declination bands. 
As a result, the dipole anisotropy can only be observed as a projection onto the equatorial plane.

\section{The Dataset}
Data selected for this analysis come from 5 years of IceCube data collected between May, 2011 and May, 2016, in its final configuration of 86 strings (IC86), 
as well as 1 year of HAWC data collected between April, 2015 and April, 2016, in its final configuration of 300 tanks (HAWC-300). Table \ref{tab:observatories} shows the characteristics of both detectors next to each other.  
Only continuous days of data containing gaps of less than 20 min were chosen for these analyses in order to reduce the bias of uneven exposure along right ascension.
There is also a difference in the median energy of the two experiments (Table \ref{tab:observatories}). The median energy grows as a function of shower zenith angle and is largest in the narrow region of overlap between the two detectors at $\delta=-40^\circ$ to $-20^\circ$."

\begin{table}[h]
\centering  
\scriptsize
\begin{tabular}{p{0.22\textwidth}|p {0.12\textwidth} | p{0.1\textwidth} | p{0.1\textwidth} | p{0.12\textwidth}}
 &\multicolumn{2}{|l}{\small \bf{IceCube}} & \multicolumn{2}{|l}{\small \bf {HAWC}} \\  \hline
Hemisphere & \multicolumn{2}{|l}{Southern}  & \multicolumn{2}{|l}{Northern}  \\
Latitude& \multicolumn{2}{|l}{-90$^\circ$} &  \multicolumn{2}{|l}{19$^\circ $}\\
Detection method &  \multicolumn{2}{|l}{muons produced by CR}  &  \multicolumn{2}{|l}{air showers produced by CR and $\gamma$}  \\
Field of view  &  \multicolumn{2}{|l}{-90$^\circ$/-20$^\circ$, ${\sim}$4 sr (same sky over 24h)} &  \multicolumn{2}{|l}{-30$^\circ$/64$^\circ$,  ${\sim}$2 sr (8 sr observed)/24 h}\\
Livetime & \multicolumn{2}{|l}{5 years}  &  \multicolumn{2}{|l}{269 days over a period of 336.36 days} \\
Detector trigger rate & \multicolumn{2}{|l}{2.5 kHz} &  \multicolumn{2}{|l}{25 kHz}   \\ \hline
 & Quality cuts & Energy cuts & Quality cuts & Energy cuts \\ \hline
Median primary energy & $20$ TeV & $10$ TeV &  $2$ TeV & $10$ TeV \\
Approx. angular resolution &  $2^\circ - 3^\circ$ & $2^\circ - 6^\circ$ &  $0.3^\circ - 1.5^\circ$ & $0.3^\circ - 1.5^\circ$ \\
Events & $2.8 \times 10^{11}$ & $1.7 \times 10^{11}$ & $2.6  \times 10^{10}$  & $4.4 \times 10^{9}$
\end{tabular}
\caption [Comparison of observatories] {
\small Comparison of the IceCube and HAWC datasets.
} \label{tab:observatories}
\end{table}
\begin{figure}[h]
        \centering
        \begin{subfigure}[b]{0.55\textwidth}
	\includegraphics[width=\textwidth]{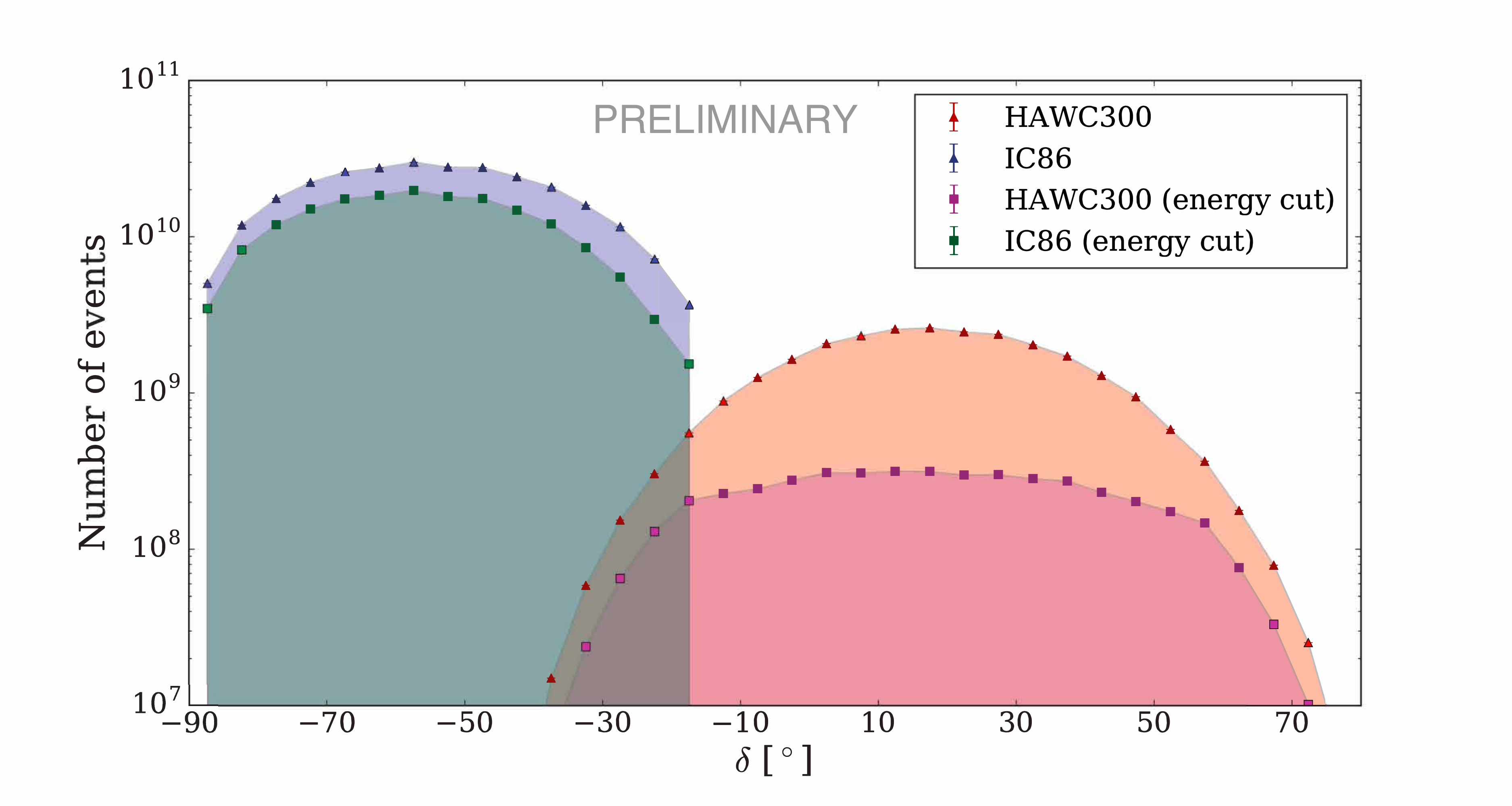} 
	\caption[Detector acceptance]{\small
	}\label{fig:detector_acceptance} 
	 \end{subfigure}
        \begin{subfigure}[b]{0.35\textwidth}
        \includegraphics[width=\textwidth]{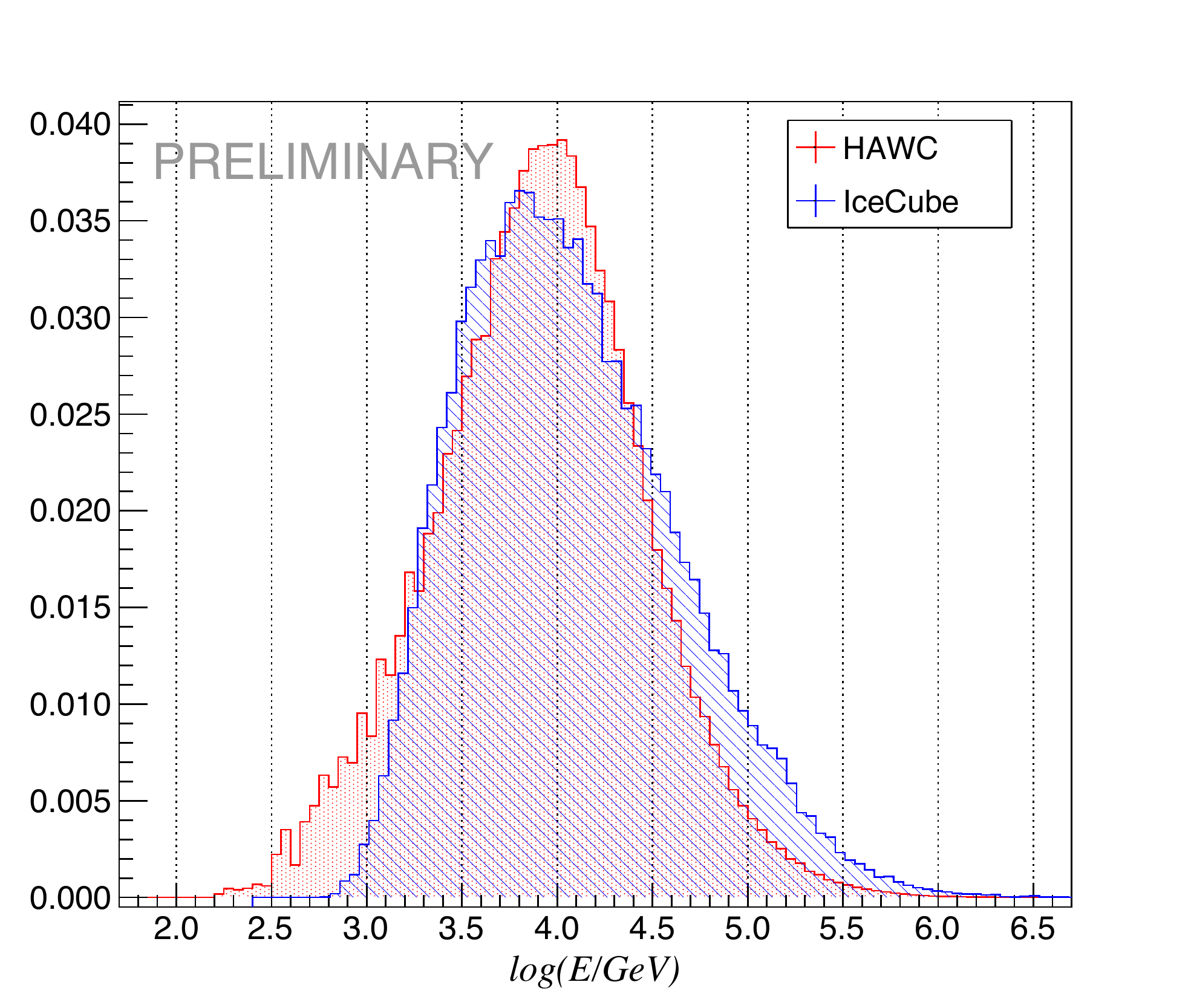}
                      \caption[Energy Distribution]{}
                \label{fig:energy_dist_sub}
        \end{subfigure}
         \vskip -0.3cm
    	\caption{\small (\subref{fig:detector_acceptance}) Distribution of events as as a function of declination for IceCube and HAWC. 
	Triangles correspond to the full energy spectrum and squares correspond to the same datasets after applying energy cuts. 
	Restricting datasets to overlapping energy bins significantly reduces statistics for HAWC. The statistics in HAWC-300 before cuts are  comparable to one year of IC86.
	(\subref{fig:energy_dist_sub}) Energy distribution in each of the two datasets after cuts.  
		 } 
        \label{fig:energy_dist}
\end{figure}

In order to select data that are consistent between the two detectors, we apply the following energy cuts: 
In the case of HAWC we use an energy reconstruction based on the likelihood method described in \cite{hamphel2017} to select events with energies at or above 10 TeV. 
We also apply a hadron/gamma separator described in \cite{Abeysekara:2017mjj} in order to exclude gamma-ray candidates. In the case of IceCube 
we apply a cut in the 2D plane of number of hit channels (which act as a proxy for muon energy) and the cosine of the zenith angle, as described in \cite{IceCube:2012feb}. 
For a given number of hit channels, events at larger zenith angles are produced by cosmic-ray particles with higher energy.
Figure \ref{fig:detector_acceptance} shows the distribution of data as a function of declination. 
The resulting energy distribution of the two datasets is shown in Figure~\ref{fig:energy_dist_sub}. 
After cuts, both CR data sets have a median energy of approximately 10 TeV with little dependence on zenith angle.

\section{Analysis}
We compute the relative intensity as a function of equatorial
coordinates ($\alpha$,$\delta$) by binning the sky into an
equal-area grid with a bin size of 0.9$^\circ$ using the 
HEALPix library~\cite{Gorski:2005apr}. The relative intensity gives the amplitude of deviations from the isotropic
expectation in each angular bin $i$. 
In order to produce residual maps of the anisotropy of the arrival
directions of the cosmic rays, we must have a description of the arrival direction distribution if the
cosmic rays arrived isotropically at Earth, $\langle N \rangle (\alpha,\delta)_i$. 
Ground-based experiments observe CRs indirectly by detecting the secondary air shower particles produced by collisions of the cosmic-ray primary in the atmosphere. Simulations are not
sufficiently accurate to describe the detector exposure at the level of $10^{-3}$. We therefore calculate this expected flux from the data themselves in order to account for rate variations 
in both time and viewing angle.

A common approach is to estimate the relative intensity and detector exposure simultaneously using time-integration methods such as {\it time-scrambling}~\cite{Alexandreas1993} and {\it direct-integration}~\cite{Atkins:2003ep}. However, these methods can lead to an under- or overestimation of the isotropic reference level for detectors located at middle latitudes, because a fixed position on the celestial sphere is only observable over a relatively short period every day. The total number of cosmic ray events from this fixed position can only be compared against reference data observed during the same period. Therefore, time-integration methods can strongly attenuate large-scale structures exceeding the size of the instantaneous field of view~\cite{0004-637X-823-1-10}. In the next section we describe a likelihood method that can recover the full amplitude of the large-scale anisotropies projected on to the equatorial plane.
%
\subsection{Maximum likelihood method} \label{sec:lh}
We apply the likelihood-based reconstruction developed by Ahlers {\it et al.} \cite{0004-637X-823-1-10}, 
generalized for combined anisotropy studies of data sets from multiple observatories that are exposed to overlapping regions of the sky. 
This likelihood method can disentangle the anisotropy from detector effects and gives a better estimate of the relative intensity of the sidereal cosmic-ray anisotropy for detectors in the middle latitudes (such as HAWC).
The cosmic ray flux can then be expressed as $\phi(\alpha,\delta) = \phi^{\rm iso}I(\alpha,\delta)$, 
where $I(\alpha,\delta)$ is the {\it relative intensity} as a function of position in the sky and $\phi^{\rm iso}$ is the isotropic angular-averaged flux. Given that cosmic rays have been observed to be mainly isotropic, 
the flux is dominated by the isotropic term and therefore the {\it anisotropy} $\delta I = I-1$ is small.

For each observatory, the number of cosmic rays expected from 
an angular element of the local coordinate sphere $\Delta\Omega_i$ corresponding to coordinates $(\theta_i,\varphi_i)$ in a sidereal time interval $\Delta t_\tau$ is
\begin{equation}\label{eq:mu}
  \mu_{\tau i} \simeq I_{\tau i}\mathcal{N}_\tau\mathcal{A}_{i}\,,
\end{equation}
where $\mathcal{N}_\tau$ gives the expected number of isotropic events in sidereal time bin $\tau$ independent of pixel,  
$\mathcal{A}_i$ is the relative acceptance of the detector for pixel $i$, and $I_{\tau i}$ is the 
relative intensity observed in the local coordinates during time bin $\tau$. The likelihood of observing $n$ cosmic rays is then given by the product of Poisson probabilities
\begin{equation}\label{eq:LH}
  \mathcal{L}(n|I,\mathcal{N},\mathcal{A}) =
  \prod_{\tau i}\frac{(\mu_{\tau i})^{n_{\tau i}}e^{-\mu_{\tau i}}}{n_{\tau i}!}\,,
\end{equation}
where $n_{\tau i}$ is the number of events observed in the local pixel $i$ during time bin $\tau$. We maximize the likelihood ratio
of signal over null hypothesis in $\mathcal{N}$, $\mathcal{A}$, and $I$,
\begin{equation}\label{eq:LHratio}
  \lambda = \frac{\mathcal{L}(n|I,\mathcal{N},\mathcal{A})}
                 {\mathcal{L}(n|I^{(0)},\mathcal{N}^{(0)},\mathcal{A}^{(0)})} \, .
\end{equation}
The maximum
$(I^\star,\mathcal{N}^\star,\mathcal{A}^\star)$ of the likelihood ratio (Eq. \ref{eq:LHratio}) results in a nonlinear set of equations that cannot be solved in an explicit form, 
but one can iteratively approach the best-fit solution.

In this combined analysis of HAWC and IceCube data, the likelihood (Eq. \ref{eq:LH}) is generalized to a product over data sets with individual detector exposures 
but the {\it same} relative intensity. 
This is a valid approach, as long as the rigidity distributions of the data sets are very similar. 
Systematic uncertainties caused by differences in energy and composition between the two data sets are currently being evaluated.
Our reconstruction method is a simple generalization of the iterative method outlined in Ahlers {\it et al.} \cite{0004-637X-823-1-10}, where now the relative acceptances $\mathcal{A}$ and isotropic expectation $\mathcal{N}$ for each detector are evaluated as independent quantities.

\subsection{Application of Method to Combined Dataset}

\begin{figure}[h]
\begin{center}
\includegraphics*[width=.60\textwidth]{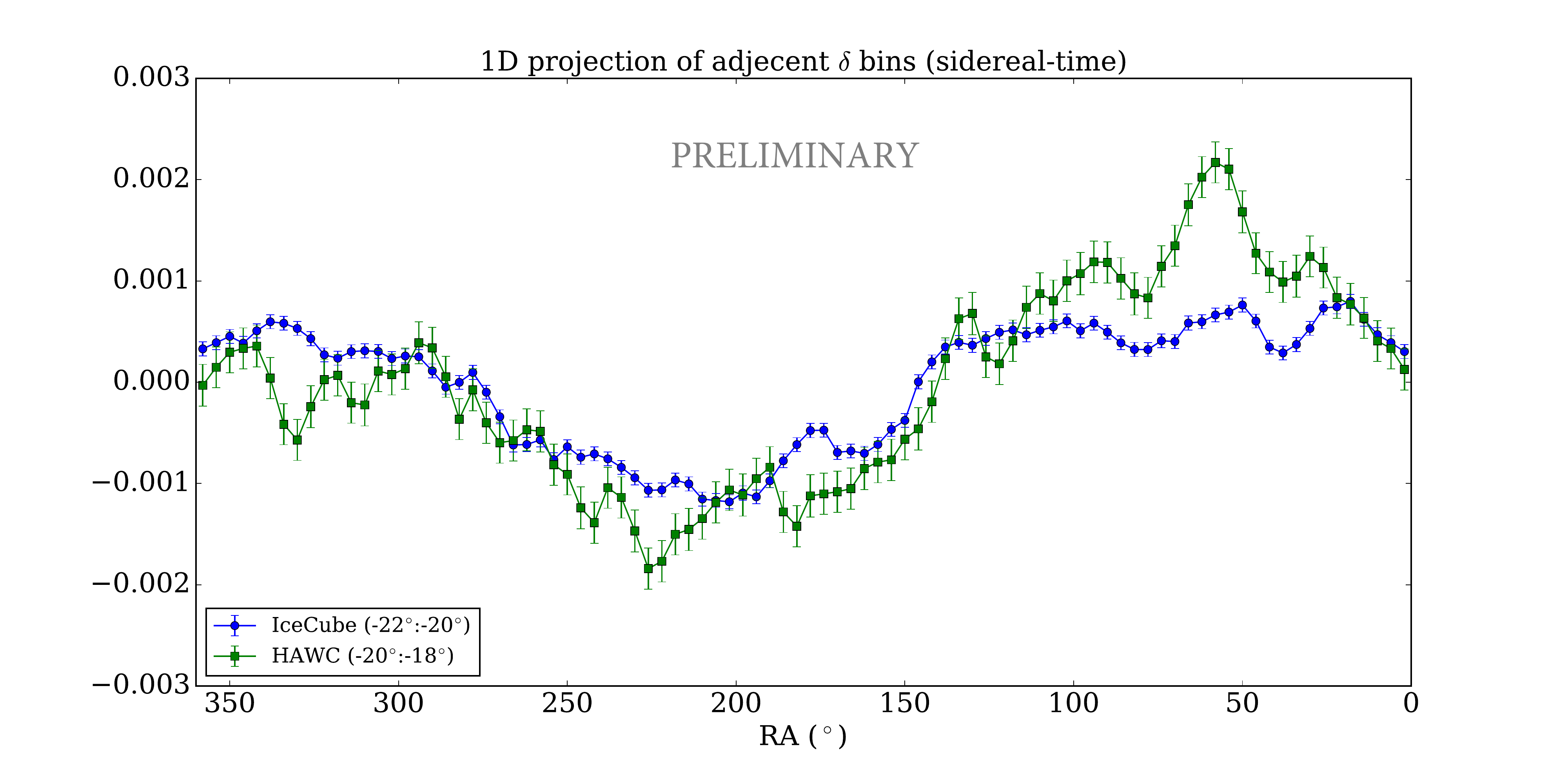} \end{center}
\vskip -0.7cm
\caption[1D projection of large-scale anisotropy.]{\small
One-dimensional RA projection of the relative intensity of cosmic rays 
for adjacent $\delta$ bins at -20$^\circ$
for HAWC-300 and IC86 data. 
There is general agreement for large scale structures. 
The two curves correspond to different $\delta$ bands but some differences in the small scale structure
might also be attributed to mis-reconstructed events that migrate from nearby $\delta$ bins with larger statistics. 
The errors are only statistical and don't account for systematic uncertainties.
} \label{fig:sidereal}
\end{figure} 
Figure \ref{fig:sidereal} shows the one-dimensional RA projection of the relative intensity of cosmic rays 
for adjacent $\delta$ bins at -20$^\circ$
for HAWC-300 and IC86 data. 
There is no overlapping region where the quality of the data is sufficient to allow a meaningful comparison so we have chosen two adjacent bins.
The large structure between the two datasets is consistent though small structures differ. 
While HAWC data has a smaller point spread function and is sensitive to structures on smaller scales,  IceCube has better statistics so the structures are more significant. 
One particular feature that stands out is the excess in HAWC around $\alpha=50^\circ$ that coincides with the so called \emph{region A}. 
There appears to be a corresponding small excess in the IceCube data. 
It is worth noting that statistics in this region are quickly decreasing with increasing zenith angle, towards the horizon of each detector, as is the quality of angular reconstructions. 
\begin{figure}[ht]
      \begin{minipage}{\textwidth} 
      \begin{center}
        \begin{subfigure}[b]{0.38\textwidth}
        \includegraphics[width=\textwidth]{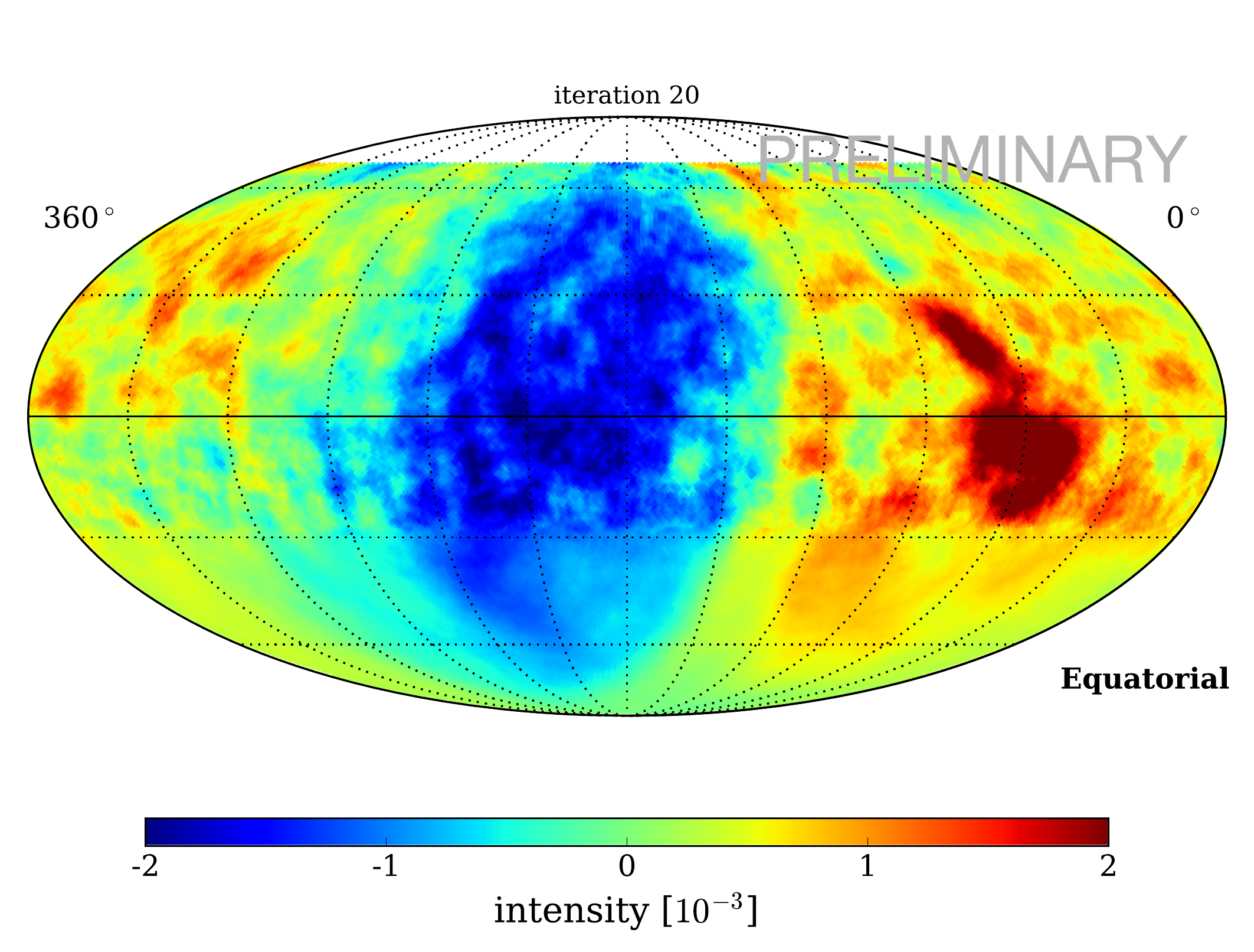}
                      \caption{}
                \label{fig:map:ls-comb}
        \end{subfigure}
        \begin{subfigure}[b]{0.38\textwidth}
        \includegraphics[width=\textwidth]{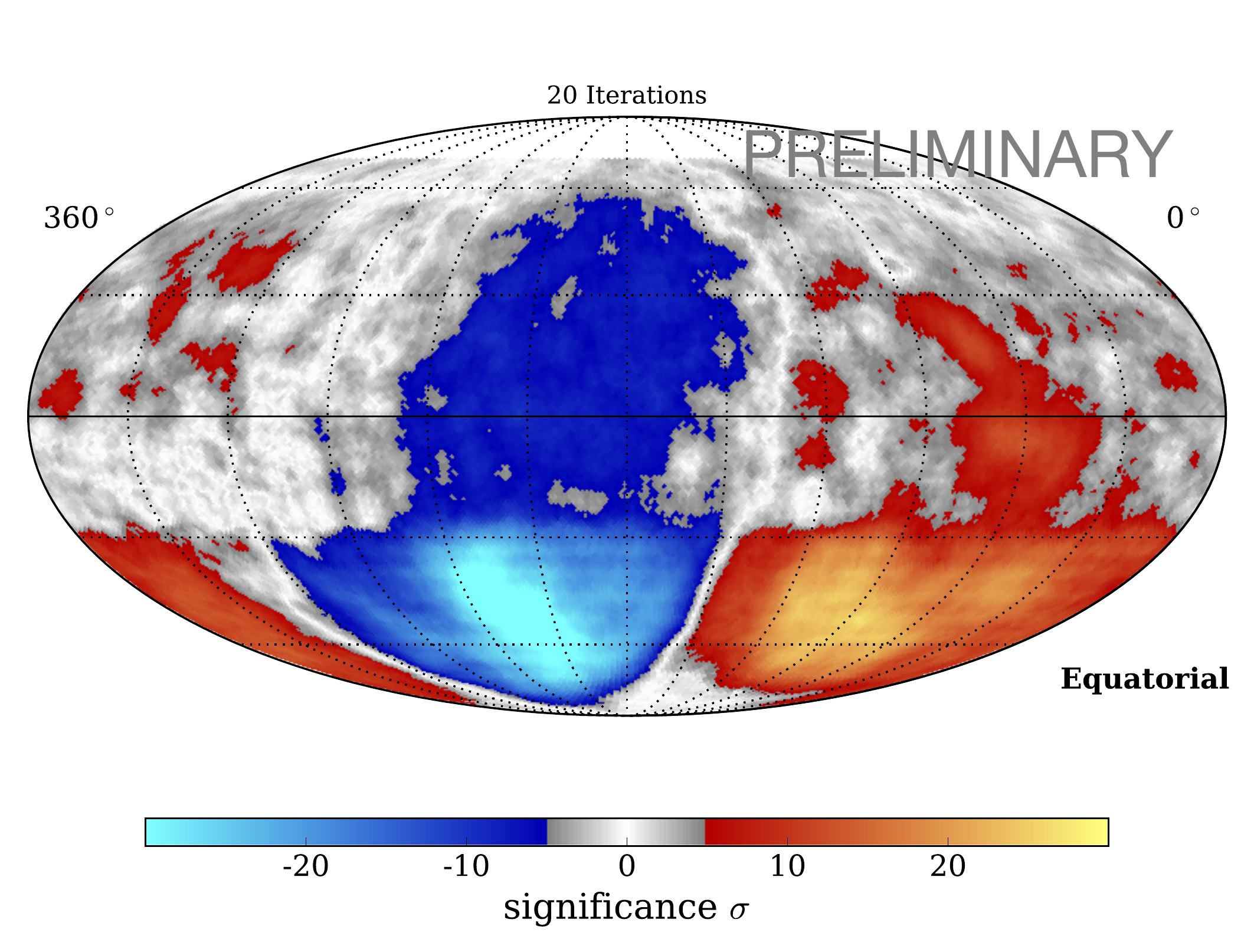}
                      \caption{}
                \label{fig:map:ls-sig}
        \end{subfigure}
        \end{center}
         \end{minipage}    \vskip -0.3cm
         \begin{minipage}{\textwidth}
            \begin{center}
        \begin{subfigure}[b]{0.36\textwidth}
        \includegraphics[width=\textwidth]{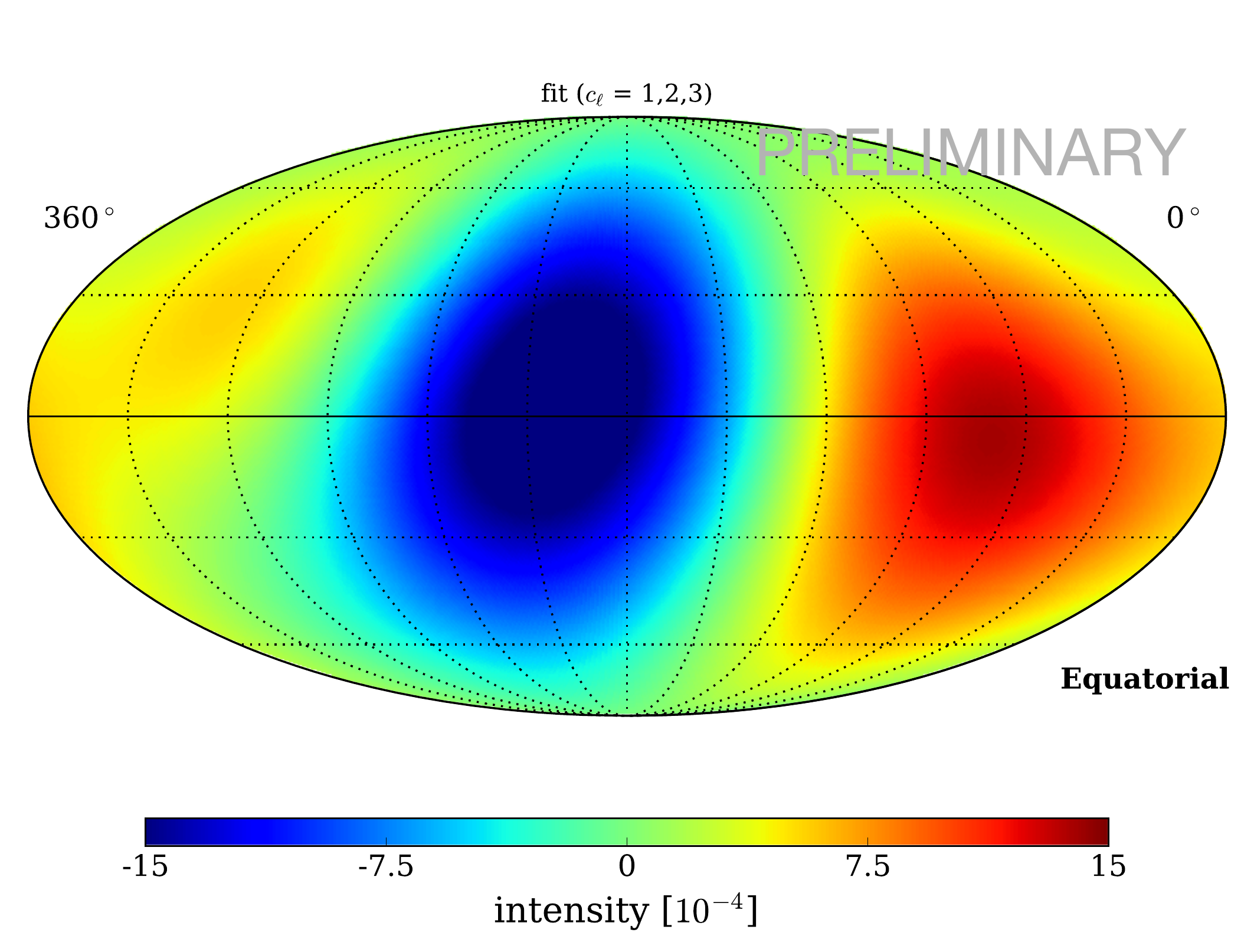}
                      \caption{}
                \label{fig:multipolefit:oct-fit}
        \end{subfigure}
        \end{center}
         \end{minipage}   \vskip -0.2cm
         \begin{minipage}{\textwidth}
               \begin{center}
        \begin{subfigure}[b]{0.38\textwidth}
        \includegraphics[width=\textwidth]{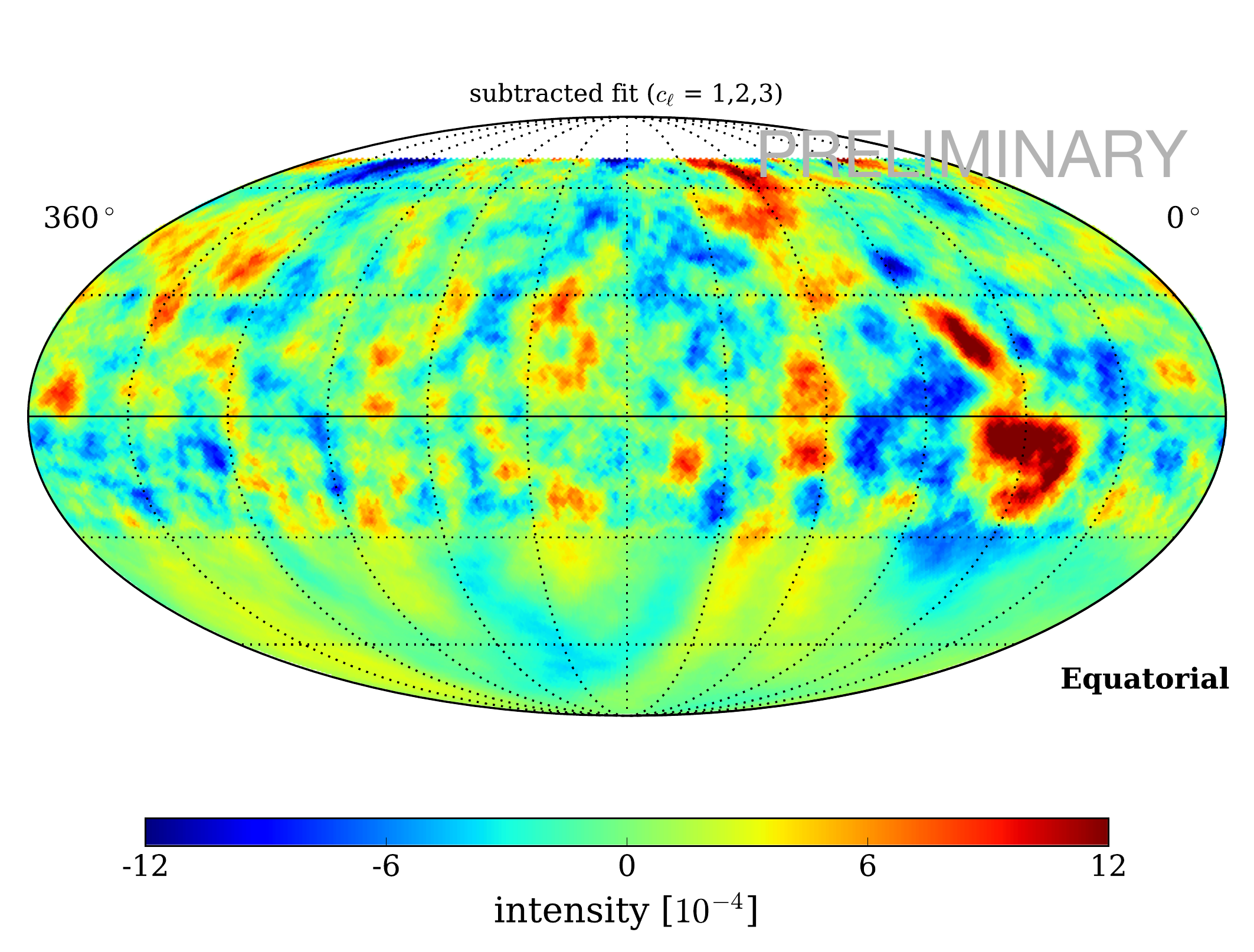}
                      \caption{}
                \label{fig:small:relint}
        \end{subfigure}
        \begin{subfigure}[b]{0.38\textwidth}
        \includegraphics[width=\textwidth]{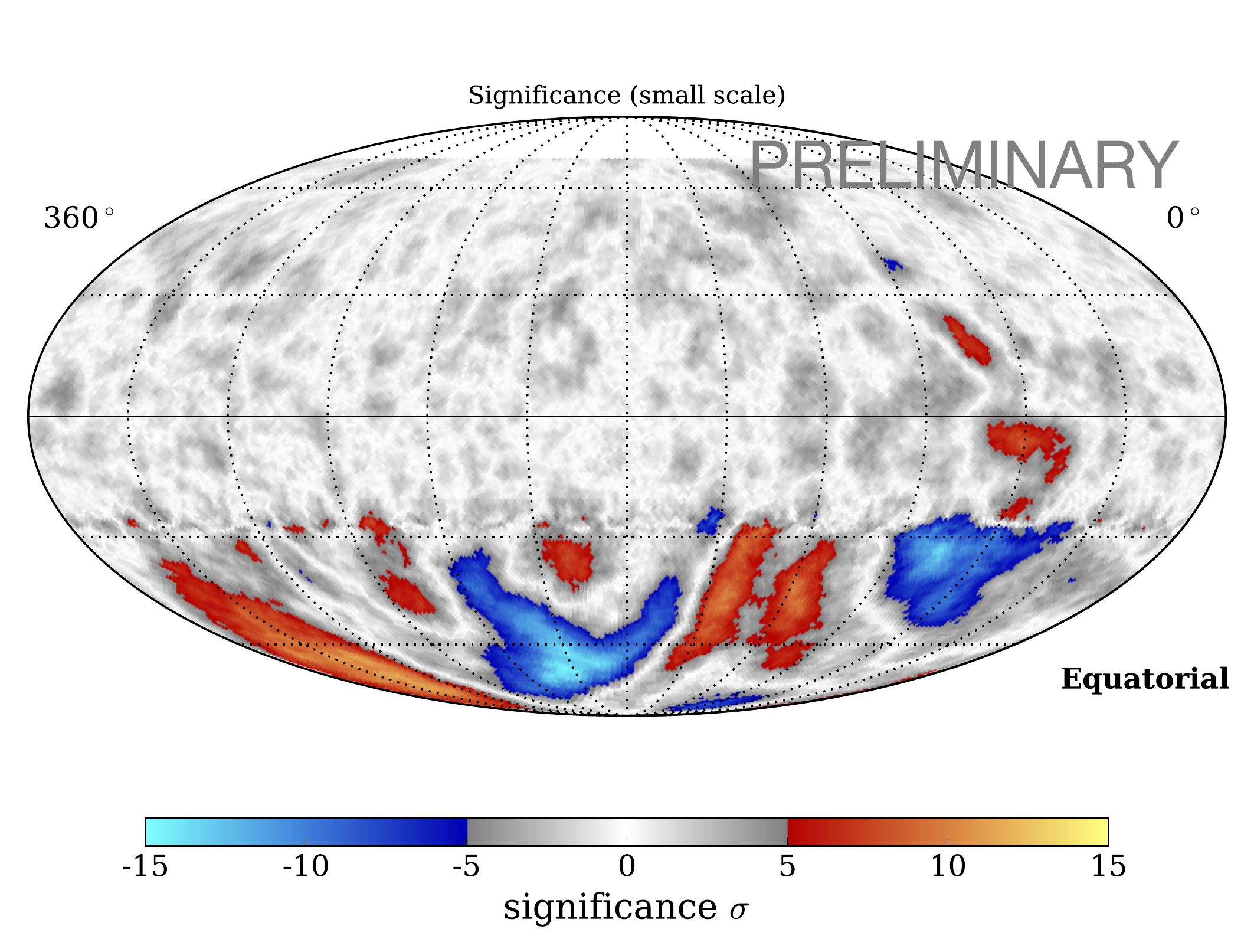}
                      \caption{}
                \label{fig:small:sig}
        \end{subfigure}
         \end{center}
         \end{minipage}
    \vskip -0.4cm
 	\caption[Combined large-scale anisotropy]{\small (\subref{fig:map:ls-comb}) Relative intensity from the maximum likelihood method after 20 iterations,  
	(\subref{fig:map:ls-sig}) corresponding statistical significance, 
		 and (\subref{fig:multipolefit:oct-fit}) multipole fit with $\ell = \{1,2,3\}$ for the combined IC86 and HAWC-300 dataset. 
		 The significance of the IceCube region reflects the much larger statistics available in 5 years of IceCube compared to 1 year of HAWC-300 at energies of $\sim$10 TeV. 
		  (\subref{fig:small:relint}) Relative intensity after subtracting the multipole fit from the large-scale map, and  (\subref{fig:small:sig}) the corresponding statistical significance.
		 } 
        \label{fig:maps}
\end{figure}

The relative intensity map obtained after 20 iterations of the 
maximum likelihood method is shown in Figure \ref{fig:map:ls-comb}. 
The corresponding statistical significance is obtained using a generalized form of the method of \cite{LiMa:1983sep} and is shown in \ref{fig:map:ls-sig}. The significance of features in the northern sky is lower than previously published HAWC results due to decreased statistics given the energy cuts in this analysis.
A  smoothing procedure has been applied to all maps using a top-hat function in which a single pixel's value is the sum of all pixels within a 5$^\circ$ radius.
The significance of the IceCube region reflects the much larger number of statistics available in 5 years of IceCube as
compared to 1 year of HAWC-300 at energies of $\sim$10 TeV. 
In order to eliminate larger structures, fitted multipoles can be subtracted to access lower angular 
scales while preserving the maximum angular scale throughout the map.
Figure \ref{fig:multipolefit:oct-fit} shows the multipole fit with $\ell = \{1,2,3\}$ for the combined IC86 and HAWC-300 dataset. 
Figure \ref{fig:small:relint} is the residual relative intensity after subtracting the fitted multipole 
of \ref{fig:multipolefit:oct-fit} from the large-scale map in \ref{fig:map:ls-comb}, and  
\ref{fig:small:sig} is the corresponding statistical significance.

\begin{figure}[h]
        \begin{subfigure}[b]{0.48\textwidth}
        \includegraphics[width=\textwidth]{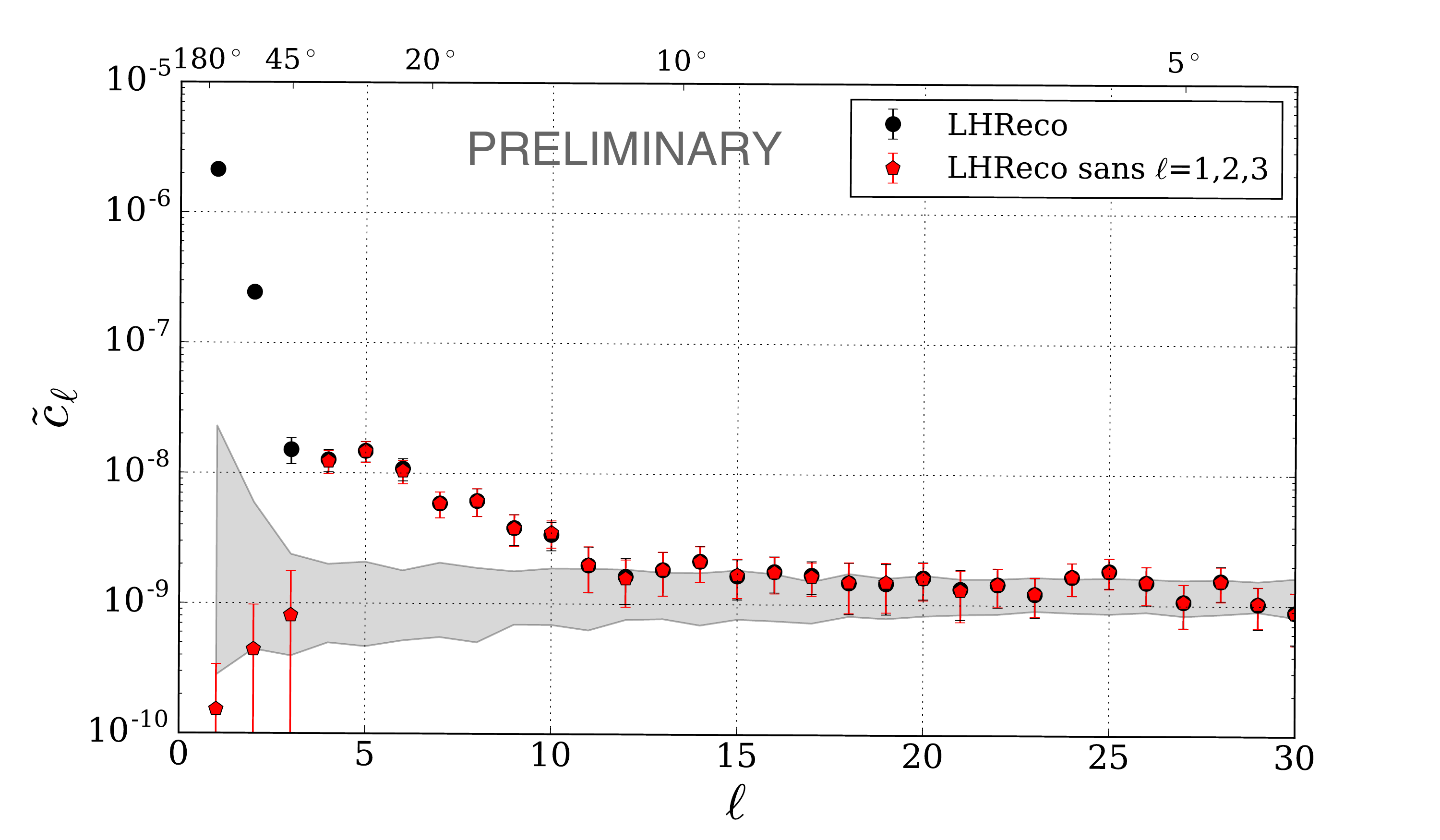}
                      \caption{}
                \label{fig:power-spectrum}
        \end{subfigure}
        \begin{subfigure}[b]{0.48\textwidth}
        \includegraphics[width=\textwidth]{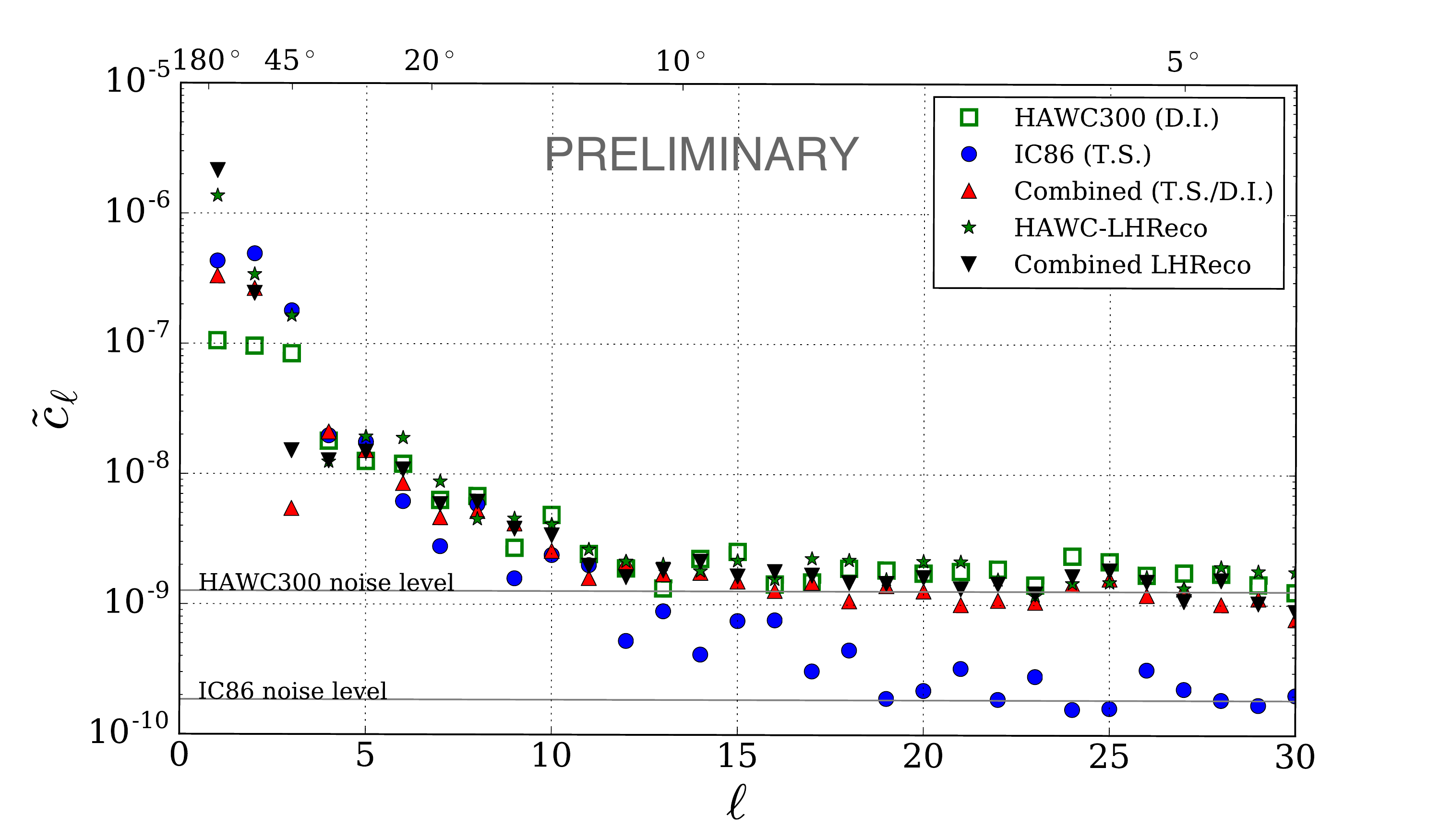}
                      \caption{}
                \label{fig:multi-spectra}
        \end{subfigure}
        \vskip -0.4cm
 	\caption[Combined large-scale anisotropy]{\small The angular power spectrum of the cosmic ray anisotropy (\subref{fig:map:ls-comb}) 
	for the combined IceCube and HAWC dataset, and
		  (\subref{fig:multi-spectra}) comparison of power-spectra for IceCube 
	and HAWC individually, and combined with time-integration and iterative methods.
	The red dots in (\subref{fig:map:ls-comb}) correspond to the power spectrum resulting from subtracting the fit to the large scale features ($\ell$ = 1, 2, 3). 
	The gray band represents the power spectra for isotropic sky maps at the 90\% confidence level.
	The large scale structure corresponds to the peak in small $\ell$ on the left while significant smaller structures up to $\ell \sim$ 10
	of angular scales between 10$^\circ$ and 35$^\circ$ can be seen above the gray band.
	The noise level in the combined dataset is dominated by limited statistics for the portion of the sky observed by HAWC.
		 } 
         \label{fig:powers-spectra}
\end{figure}

\begin{figure}[h]
 \centering
\includegraphics[width=.52\textwidth]{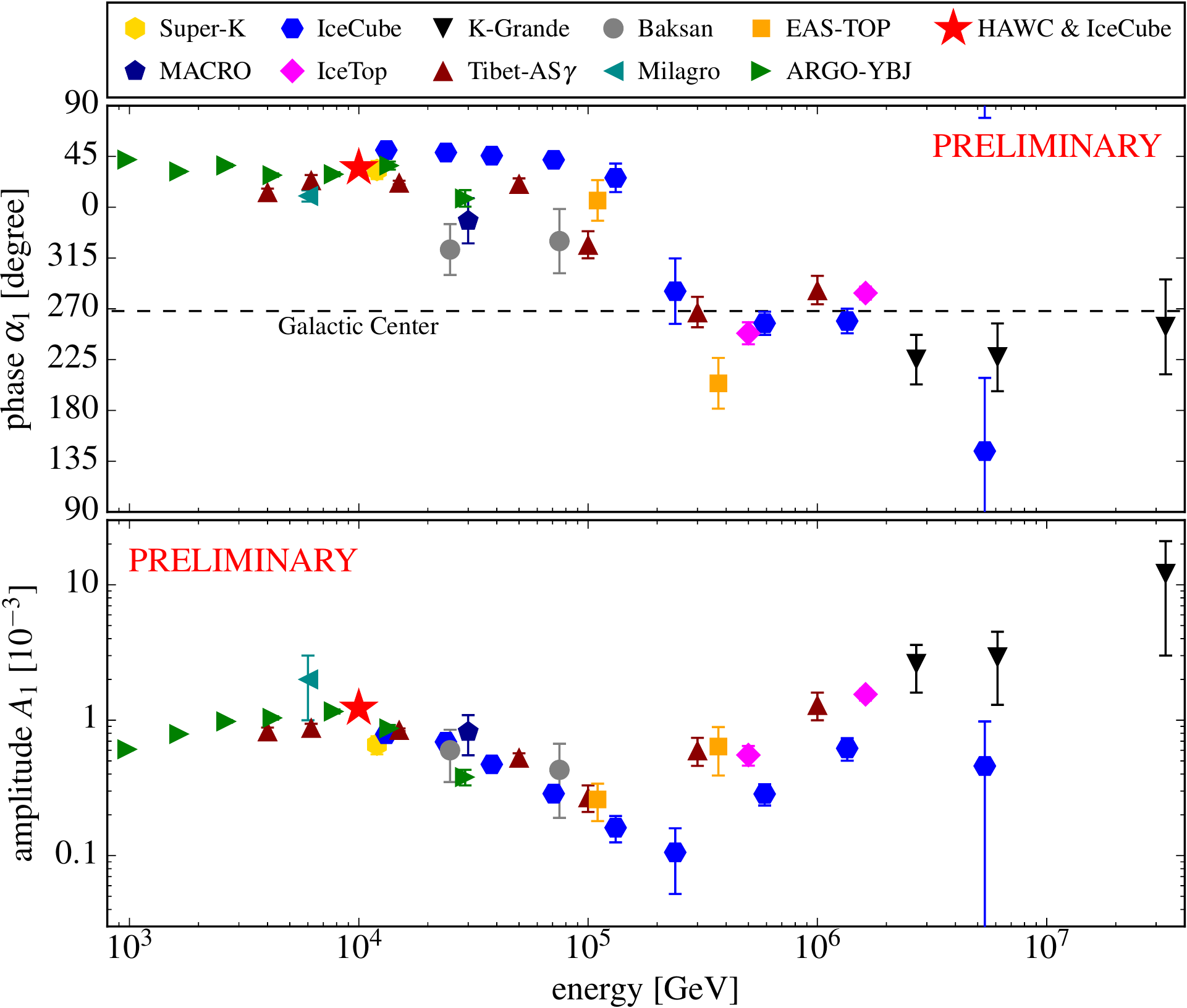}
 \vskip -0.2cm
\caption[]{\small
Summary plot (adopted from Ref.~\cite{Ahlers:2016rox}) of the reconstructed TeV-PeV dipole amplitude and phase (see Ref.~\cite{Ahlers:2016rox} for details and references).}\label{fig:phase}
\end{figure}

Figure \ref{fig:power-spectrum} shows the angular power spectrum of the resulting relative intensity of cosmic rays 
for the combined IceCube and HAWC dataset calculated with the method described in \cite{IceCube:2012feb} and \cite{HAWC:2014dec}.
The large scale structure corresponds to the peak in small $\ell$ while significant smaller structures up to $\ell \sim$ 10
of angular scales between 10$^\circ$ and 35$^\circ$ can be seen above the gray band.
The red dots correspond to the power spectrum resulting from subtracting the fitted multipoles ($\ell$ = 1, 2, 3) and removing the large scale features. 
In Figure \ref{fig:multi-spectra} the likelihood method (stars) reconstructs a stronger dipole ($\ell=1$)  component compared to using direct integration over 24h (squares).
The full-sky coverage also provides better constraints for fitting the quadrupole ($\ell = 2$) and octupole ($\ell = 3$) components and reduces cross-talk between spherical harmonic expansion coefficients $a_{\ell m}$.
Figure \ref{fig:phase} shows the reconstructed dipole phase and amplitude 
from this analysis in the equatorial plane along with data from several other experiments (from \cite{Ahlers:2016rox}).

\section{Conclusions and Discussion}
The analysis of five years of data taken with the IceCube detector and 1 year of HAWC
shows an anisotropy in the arrival direction distribution of 10 TeV
cosmic rays that extends across both hemispheres. In this analysis we have used an 
iterative maximum-likelihood reconstruction method that simultaneously fits cosmic ray anisotropies and detector
acceptance. 
The method does not rely on detector simulations and provides an optimal anisotropy
reconstruction and the recovery of the dipole anisotropy for ground-based cosmic ray observatories located in the middle latitudes such as HAWC.
Ground-based observatories are generally insensitive to cosmic-ray anisotropy variations that are symmetric in right ascension, i.e. only vary across declination bands.
In particular, the dipole anisotropy can only be observed as a projection onto the celestial equator.
The combined dataset provides almost full coverage of the sky and provides a better fit for the phase and amplitude 
of the horizontal component of the dipole anisotropy.
In addition to a large-scale structure, we observe significant
small-scale structure that is largely consistent with previous individual measurements. 

\bibliographystyle{ICRC}
\bibliography{IC86_HAWC_CR_Aniso}

\end{document}